\documentclass{article}

\usepackage{arxiv}

\usepackage[utf8]{inputenc} 
\usepackage[T1]{fontenc}    
\usepackage{hyperref}       
\usepackage{url}            
\usepackage{booktabs}       
\usepackage{amsfonts}       
\usepackage{nicefrac}       
\usepackage{microtype}      
\usepackage{cleveref}       
\usepackage{lipsum}         
\usepackage{graphicx}
\usepackage{natbib}
\usepackage{doi}

\title{Exploring the interplay of individual traits and interaction dynamics in preschool social networks}

\date{June 28, 2024}

\usepackage{authblk}

\newbox{\orcid}\sbox{\orcid}{\includegraphics[scale=0.06]{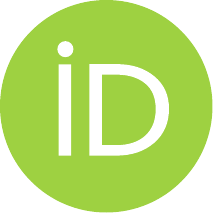}} 

\author[1]{%
	\href{https://orcid.org/0000-0001-9572-9352}{\usebox{\orcid}\hspace{1mm}Gülşah Akçakır\thanks{\texttt{gakcakir@ucla.edu}}}%
}
\author[2]{%
	\href{https://orcid.org/0009-0008-9438-2527}{\usebox{\orcid}\hspace{1mm}Amina Azaiez\thanks{\texttt{amina.azaiez@univ-paris1.fr}}}%
}
\author[3]{%
	\href{https://orcid.org/0000-0002-1423-249X}{\usebox{\orcid}\hspace{1mm}Alberto Ceria\thanks{\texttt{a.ceria@liacs.leidenuniv.nl}}}%
}
\author[4]{%
	\href{https://orcid.org/0000-0002-2686-4814}{\usebox{\orcid}\hspace{1mm}Clara Eminente\thanks{\texttt{eminente\_clara@phd.ceu.edu}}}%
}
\author[5]{%
	\href{https://orcid.org/0009-0005-1862-9856}{\usebox{\orcid}\hspace{1mm}Guglielmo Ferranti\thanks{\texttt{guglielmo.ferranti@phd.unict.it}}}%
}
\author[6]{%
	\href{https://orcid.org/0000-0002-2398-7480}{\usebox{\orcid}\hspace{1mm}Govind Gandhi\thanks{\texttt{gogandhi@iu.edu}}}%
}
\author[7]{%
	Aishvarya Raj \thanks{\texttt{aishvarya.raj@lis.ac.uk}}%
}
\author[8,9]{%
	\href{https://orcid.org/0000-0001-8794-6410}{\usebox{\orcid}\hspace{1mm}Iacopo Iacopini\thanks{\texttt{iacopo.iacopini@nulondon.ac.uk}}}%
}
\affil[1]{Department of Communication, University of California, Los Angeles, CA, USA}
\affil[2]{Department of Applied Mathematics, Panthéon Sorbonne University, 75013 Paris, France}
\affil[3]{Leiden Institute of Advanced Computer Science, Leiden University, Leiden, The Netherlands}
\affil[4]{Department of Network and Data Science, Central European University, 1100 Vienna, Austria}
\affil[5]{Department of Physics and Astronomy “Ettore Majorana”, University of Catania, 95123 Catania, Italy}
\affil[6]{Department of Informatics, Indiana University Bloomington, IN, USA}
\affil[7]{The London Interdisciplinary School, London, United Kingdom}
\affil[8]{Network Science Institute, Northeastern University London, E1W 1LP London, United Kingdom}
\affil[9]{Department of Physics, Northeastern University, Boston, MA 02115, USA}

\hypersetup{
	pdftitle={Exploring the interplay of individual traits and interaction dynamics in preschool social networks},
	pdfauthor={Gülşah Akçakır, Amina Azaiez, Alberto Ceria, Clara Eminente, Guglielmo Ferranti, Govind Gandhi, Aishvarya Raj, Iacopo Iacopini},
}

\begin{document}
	\maketitle
	
	\begin{abstract}
		Several studies have investigated human interaction using modern tracking techniques for face-to-face encounters across various settings and age groups. However, little attention has been given to understanding how individual characteristics relate to social behavior. This is particularly important in younger age groups due to its potential effects on early childhood development. In this study, conducted during the Complexity 72h Workshop, we analyze human social interactions in a French preschool, where children's face-to-face interactions were monitored using proximity sensors over an academic year. We use metadata from parent surveys and preschool linguistic tests, covering demographic information and home habits, to examine the interplay between individual characteristics and contact patterns. Using a mixture of approaches, from random forest classifiers to network-based metrics at both dyadic and higher-order (group) levels, we identify sex, age, language scores, and number of siblings as the variables displaying the most significant associations with interaction patterns. We explore these variables' relationships to interactions within and outside classrooms and across mixed and single-grade classes. At the group level, we investigate how group affinity affects group persistence. We also find that higher-order network centrality (hypercoreness) is higher among children with siblings, indicating different group embedding despite similar total contact duration. This study aligns with existing literature on early social development and highlights the importance of integrating individual traits into the study of human interactions. Focusing on 2-5-year-olds offers insights into emerging social preferences during critical phases of cognitive development. Future research could use these findings to enhance mechanistic models of complex social systems by incorporating individual traits.
	\end{abstract}
	
	\section{Introduction}
	Complex social systems, consisting of numerous interacting individuals, can be effectively represented by networks of nodes connected via edges~\cite{albert2002statistical, newman2003structure, latora_nicosia_russo_2017}. Since many different natural and man-made systems can be characterised through individual interacting parts, complex networks have emerged as a universal representation to analyse and model them~\cite{barrat2008dynamical, vespignani2012modelling}. While the study of networks began with graph theory in the field of discrete mathematics \cite{euler1741solutio}, it has increasingly been applied to a wide range of systems. In the social sciences~\cite{wasserman1994social, castellano2009statistical}, networks have helped to understand, among other things, the emergence of norms~\cite{baronchelli2018emergence}, cooperation~\cite{axelrod1984evolution}, the dynamics of social contagion \cite{mitchell1973networks, centola2007complex}, and the formation of opinions and consensus \citep{nowak1990private, sznajd2000opinion}.
	
	Over the past decades, advancements in automated data collection technologies have enhanced researchers' ability to feed empirical data into theoretical approaches, allowing the integration of insights obtained from observational studies of social systems into mathematical modeling. In particular, sensors like Radio Frequency Identification (RFID) tags enable the collection of comprehensive real-time data on close-range or face-to-face interactions~\cite{isella2011s, barrat2014measuring, mastrandrea2015contact}. Research in the field of proxemics has uncovered correlations between social and spatial distances, where a closer physical distance between interacting individuals often indicates a more intimate relationship~\cite{cristani2011towards}. More generally, taken as a proxy for social interactions, these longitudinal data of dyadic contacts between individuals can then be studied within the framework of temporal (social) networks~\cite{holme2012temporal}, where the strength of connections can map the frequency or the duration of interactions~\cite{genois2019building}. The temporally-resolved nature of these empirical data collections also enables higher-order network representations, where dyadic interactions are replaced by groups~\cite{battiston2020networks, battiston2021physics} ---and links become hyperlinks connecting an arbitrary number of nodes~\cite{torres2021and, bick2023higher}. Such networks can subsequently be analysed in order to learn more about an individual's importance or role within the network, in particular with respect to the context of interaction and the research question behind the study. For example, dominant nodes within a network can be identified through network centrality measures. These measures can then be used to locate and analyse peripheral nodes or, in the context of social systems, isolated members of the group.
	
	Significant efforts have been put forward towards developing more realistic models of social interaction starting from data collected through these types of technologies~\cite{stehle2010dynamical, zhao2011social, perra2012activity, starnini2013modeling, vestergaard2014memory, karsai2014time, nadini2018epidemic, lebail2023modelling}, both in pairwise and non-dyadic settings~\cite{petri2018simplicial, gallo2024higher, iacopini2023temporal}. Despite the success of these data-driven modeling approaches, they often neglect the individual characteristics that can influence the observe social dynamical. The integration of heterogeneous traits at the level of nodes could instead lead to a better characterisation of the driver of social interactions, enhancing our understanding of these systems and our ability to inform realistic models. 
	
	While research on social interaction patterns often falls short of supplementing contact information with social identity markers and individual characteristics beyond basic demographics like gender or age, there are exceptions~\cite{kontro2020combining, dai2022longitudinal, genois2023combining}. In one such study~\cite{stehle_gender_2013}, investigating the contact preferences of 6–12 years old children in a primary school in France led to evidence for sex-based homophily. The observation that sex-based homophily increases with age is made more complex when considering weak ties. In this case, sex homophily decrease with grade for girls, yet increases with grade for boys \cite{stehle_gender_2013}. Furthermore, even when metadata on the participants are available, studies typically focus on one dimension at a time or, at most, interactions between two dimensions. However, relying  on multidimensional traits to assess social interactions holds promise of gaining a better understanding of dynamics of behavioral networks \cite{mastrandrea_contact_2015}. Multidimensional traits can reveal complex interdependencies and patterns that single or two-dimension analyses might overlook, providing deeper insights into how various social factors interplay and influence contact preferences at the dyadic and group level.
	
	In this study, we investigate the interplay between individual characteristics and social interactions in a French preschool~\cite{dai2022longitudinal}, utilizing time-resolved data on face-to-face proximity collected through wearable sensors over different months throughout the time-span of a year. The study of early childhood interactions provides valuable insights into how developmental factors impact the functioning of human social systems. In fact, through interaction with peers and caregivers, children between the ages of~3 and 6 develop skills in social and emotional regulation \cite{mcclelland2007links}. Moreover, research in the fields of social-emotional competence and school-readiness demonstrate that social competence of children entering kindergarten is a strong predictor for later academic success~\cite{denham2006social, birch1997teacher}. Getting a deeper understanding of social interaction in early childhood can also inform us about the development of social preferences and homophily. Several studies demonstrate that social links are more likely to be formed between individuals who share traits such as gender, race, and social status \cite{mcpherson2001birds, mayhew1995sex}. Further insights can thus be gained by examining how these phenomena emerge during the critical period when children develop social skills and are influenced by their peers.
	Controlled experiments conducted in educational institutions of a higher level with young children or adult participants (e.g., \cite{mastrandrea_contact_2015, stehle_gender_2013}) do not fully inform us about the factors contributing to formation and dynamics of behavioral networks in the early childhood period. Considering prominent differences in structure and rules both in- and out-of-class settings (e.g., seating arrangements) on top of overall developmental level, one anticipates to find substantial differences in the preferences. Primary school students and beyond typically spend the entire class time seated near peers, either by choice or through assigned seating arrangements. Preschoolers, on the other hand, are usually freer to move around and interact with a broader range of classmates even in class.
	
	Research on early childhood social dynamics utilising wireless sensor networks has gained some attention in recent years. The collection of movement and speech data with sensing technologies in pre-school settings has been used to learn about the relationship between language development and social interactions \cite{elbaum2024investigating}. In fact, movement data alone can provide a rich description of social dynamics since this data captures both information on proximity and synchrony, which are distinct indicators used distinguish between friendships and ephemeral interactions to achieve temporary goals \cite{horn2024automated,santos2015affiliative}. Findings on age-related differences in expectations of inter-personal distancing \cite{paulus2018preschool} show that children develop reason abilities about inter-personal space at an early age, supporting the use of sensor methods in pre-school classrooms. Network approaches to studying peer to peer relationships can reveal links between developmental disabilities such as autism spectrum disorder and lower social connectedness and isolation \cite{chen2019social,locke2013social}. Similar results were found by Chamberlain et al., in a study on the involvement on autistic children in typical classrooms; children with high functioning autism or autistic spectrum disorder experienced lower centrality, acceptance, companionship, and reciprocity, but not lower levels of loneliness\cite{chamberlain2007involvement}. In recent work exploring the links between socio-demographic traits and homophily in pre-school children using a range of individual and group-based indices found that children choose to interact similar others, particularly with respect to sex and linguistic development features. Additionally, sex-based homophily increased with age \cite{dai2022thesis}. Previous work on the structure of social groups has shown that tendencies towards cluster formation differs between children and adults. In particular, it was found that the levels of transitive organisation increased from the age of 3 to 11. This is said to reflect cognitive development in children giving rise to interpersonal preferences\cite{leinhardt1973development}.
	
	In the following report, we identify the relevant socio-demographic and linguistic features that exhibit variability among the pupils, and the consistent interaction patterns during both in-class and out-of-class periods. We do this by examining both dyadic and higher-order signatures of social interactions, accounting thus for the existence of social groups of different sizes that be combined in non-trivial patterns. Finally, we look for differences in interaction patterns when accounting for individual traits.
	
	\section{Materials and Methods}
	\label{sec:mat_and_methods}
	
	\subsection{Dataset}
	\label{sec:dataset}
	
	This work utilizes data from the DyLNet project \cite{dai2022longitudinal}. The goal of the original project was to observe the co-evolution of social networks and language development of children in pre-school age. The dataset comprises information on a total of 164 children and their interactions within school setting. Social interactions were estimated using spatial proximity sensors. Proximity data was collected in the span of 10 months. Specifically, recordings were made during one week in each of the 10 months and in 9 sessions per week. 
	Alongside the information on interactions, the dataset also contains the metadata collected through a survey administered to the parents and language tests administered to the children. The survey is composed of basic socio-demographic information on the children as well as additional information regarding their attitudes, favorite activities outside of school, and home environment. The language tests were designed to assess the level of language development of the participants.
	
	The authors of Ref.~\cite{dai2022thesis} already investigated the relationship between network structure and some additional features by means of homophily. In particular, they investigated whether networks aggregated over the timespan of four months showed homophilic behavior with respect to gender, dominant language of the child, occupation category of the mother, occupation category of the father, education level of the mother, education level of the father, vocabulary size, and syntactic development level. They found that the dimension showing the most significant level of homophily is gender, especially during time spent out of class. Moreover, they found that the level of homophily is very close to the baseline for all dimensions during class time, probably due to the the fact that during class-time children are assigned fixed seats.
	
	\subsection{Individual characteristics}
	\label{sec:characteristics}
	
	The uniqueness of this dataset lies in the metadata that comes attached with the individual nodes.
	Parents of the children participating in the study were asked to fill in questionnaires in order to provide a basic socio-demographic characterization of the children (gender, age) as well as other information regarding their environment at home and activity preferences. 
	Information relevant to assess the link between language and social interactions was collected through the questionnaire and a series of language tests. Parents were asked to indicate the main language spoken in the household and whether the children could understand and speak a language other than French. Additionally, children were administered individual language tests in order to evaluate their receptive lexical skills, short-term memory and their receptive syntactic skills, and the individual scores from these tests were recorded in the dataset.  
	
	\begin{figure}
		\centering
		\includegraphics[width=0.8\textwidth]{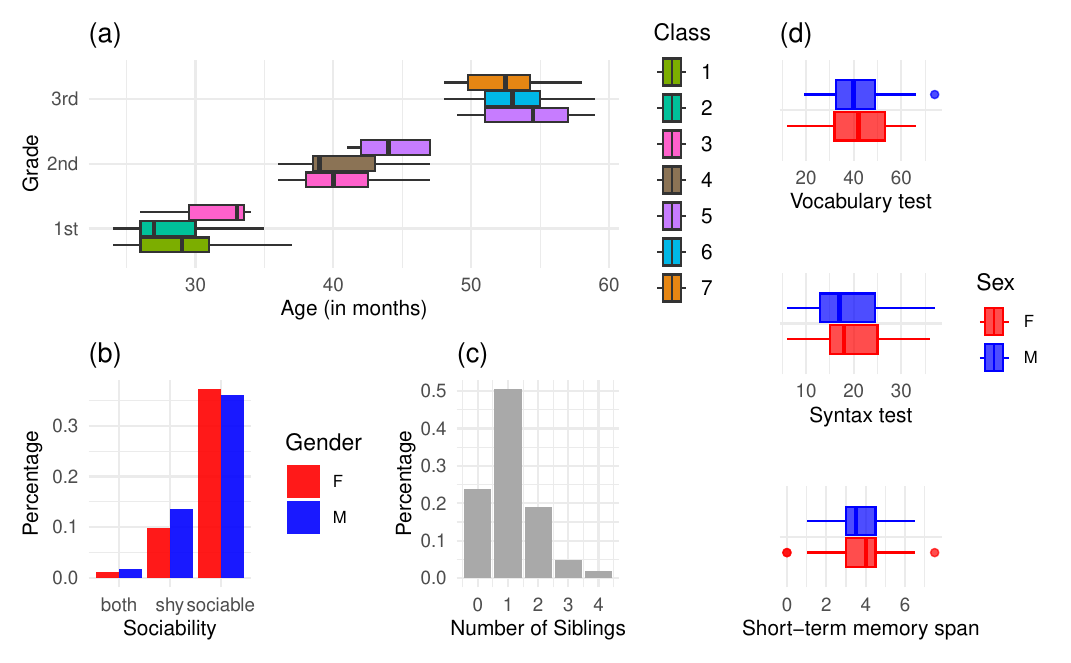}
		\caption{Descriptive statistics of individual socio-demographic and linguistic features. (a) Distributions of childrens' age by their grade and class they were assigned to. Notice the presence of classes with mixed grades. (b) Percentage of answers given by the parents when asked is their child rather sociable or shy, disaggregated by sex. (c) Distribution of number of siblings that the children have. (d) Distributions of aggregated scores in development skill tests disaggregated by sex.}
		\label{fig:combMeta}
	\end{figure}
	
	We first inspected the descriptive statistics of the survey data to identify features that can augment our understanding of social behavioural patterns when combined with the high-resolution contact data. Among the participating children, the distribution of sex is comparable, with 81 female and 83 male. The age gap spans nearly three years, ranging from a minimum of 24 months to a maximum of 59 months. Based on their school level (i.e., $1^{st}$, $2^{nd}$ and, $3^{rd}$ grade), children were divided into 7 classes. Two classes have mixed grades (3 and 5), with one combining $1^{st}$- and $2^{nd}$-graders, and the other combining $2^{nd}$- and $3^{rd}$-graders. Figure \ref{fig:combMeta}(a) illustrates the age and grade composition of each class. As expected, the mixed classes show the highest age variability, with ages ranging from 26 to 44 months in Class 3 and 41 to 57 months in Class 5. Overall, we observe that sex balance was maintained across the classes, except for one class where there was a 70-30\% split, with male students in the majority. 
	
	The survey questions regarding children's personality traits were restricted to only two aspects: sociability and talkativeness. Figure~\ref{fig:combMeta}(b) demonstrates the distribution of parent responses to the question of if the child is social or shy, categorized by sex. We observe that parents perceive a higher rate of male pupils as shy compared to females, whereas females are more likely to be categorized as sociable. The information on talkativeness lacks variability and so is not included in the further analyses. As illustrated in Figure \ref{fig:combMeta}(c), the number of siblings is another variable that exhibits variation among the group of pupils who participated in the original study, with the majority having at least one sibling.
	
	The dataset also includes two indicators per both lexical and syntactic skills, as well as short-term memory span evaluations administered at two different time points, resulting in a total of 10 measures. The linguistic evaluations include test items specifically designed for the level of each grade and 10 anchor questions presented to the children whichever their grade ---and chosen to be rather adapted to 3rd grade pupils. Since the test items measuring the same type of skills are highly correlated (not shown), we aggregated them by averaging, for each skill, the sum of distinct scores from the two consecutive years. These two resulting aggregated linguistic scores show a correlation, yet we still observe considerable variation across different score ranges. Conversely, short-term memory span does not correlate with linguistic skills. Probably unsurprisingly, these developmental measures, even though separately designed for each grade, turn out to be correlated with age, but at moderate levels. The boxplots for the three resulting development skill test measures are reported in Figure \ref{fig:combMeta}(d), disaggregated by sex.
	
	\begin{figure}
		\centering
		\includegraphics[width=\textwidth]{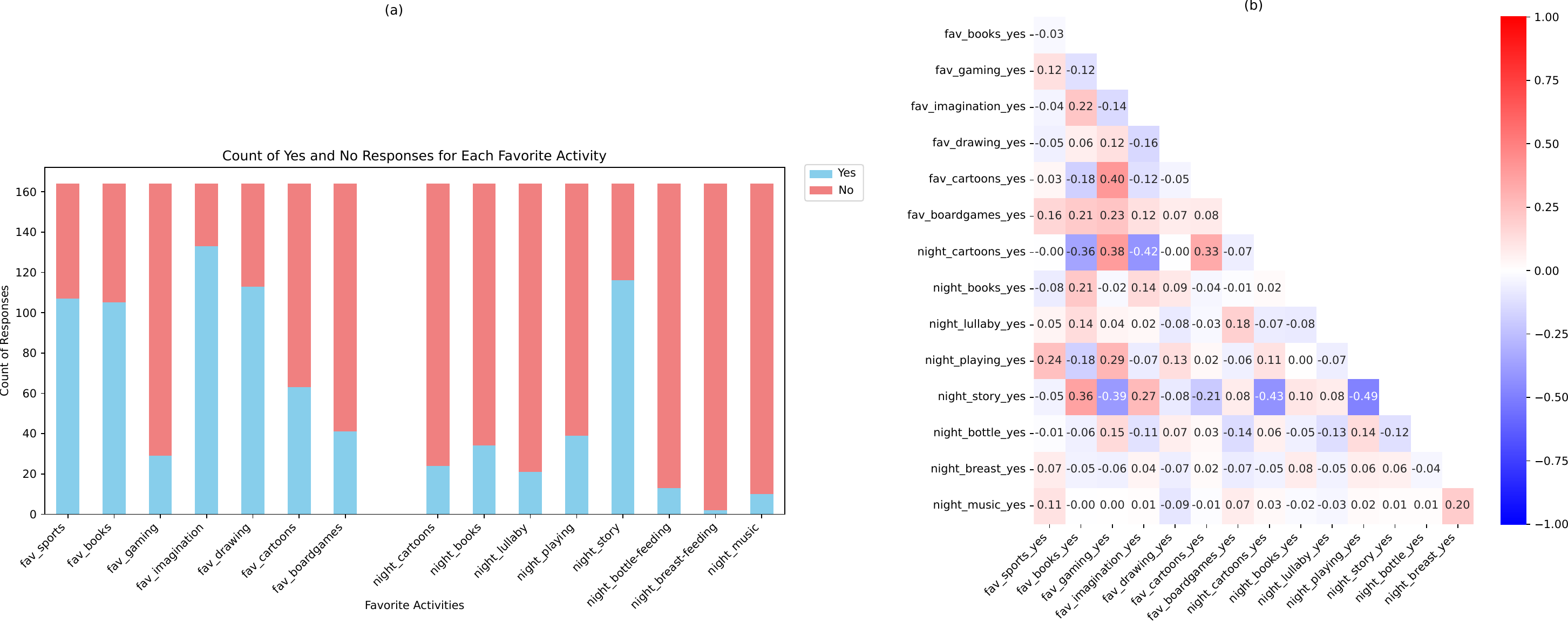}
		\caption{Descriptive analysis of preferred home activities among individuals. (a) Number of positive and negative responses associated to the different out-of-school activities during day (leftmost bars) and before nightime (rightmost bars). (b) Correlation matrix of favorite activities.}
		\label{fig:fav_activities}
	\end{figure}
	
	Parents were also asked to indicate their children's preferred activities from a set of proposed ones, distinguishing between daytime and nighttime activities. Aggregated counters associated to the given answers are reported in Figure \ref{fig:fav_activities}(a), while the correlations between them is displayed in Figure \ref{fig:fav_activities}(b). We first notice that the sample of children is quite heterogeneous, and most of the variables are poorly correlated. Additionally, we also see from (a) that some variables exhibit low variability and may not effectively distinguish social behaviors. Nevertheless, we decided to retain all these features to avoid arbitrary selections, while we did not use dimensionality reduction techniques to preserve their individual interpretation.

	\subsection{From proximity data to networks}
	\label{sec:networks}
	
	Information on spatial proximity and contact duration was collected using autonomous RFID Wireless sensors installed on participants. These devices tracked face-to-face proximity with a resolution of 5$s$. The experimental set-up included in-situ checks and controlled experiments, in order to validate the information recorded through the sensors (i.e. making sure it corresponds to non spurious interactions). More details on the validation pipeline are given in Ref.~\cite{dai2020temporal}. Once validated, the proximity data was aggregated in mutually observed pairs to create an undirected network. For the scope of this study, the data was further aggregated with a time resolution of 10$s$, associated to a proximity of participants within $~2$ meters~\cite{iacopini2023temporal}. Additional checks were performed to be sure that this choice did not affect the fundamental network properties of the resulting contact network.
	
	The obtained network was then used to investigate the social interaction patterns among pupils. We assigned to each link, e.g. the one between nodes/children $i$ and $j$, a corresponding weight $w_{ij}$, accounting for the total amount of time spent together. The individual propensity of a node $i$ to interact with its peers can be quantified via the node strength $s_i$, which is the total amount of time spent interacting with any other node. Finally, the number of different individuals that a node $i$ interacts with is quantified by its degree, $k_i$.
	
	\begin{figure}
		\centering
		\includegraphics[width = 0.8\textwidth]{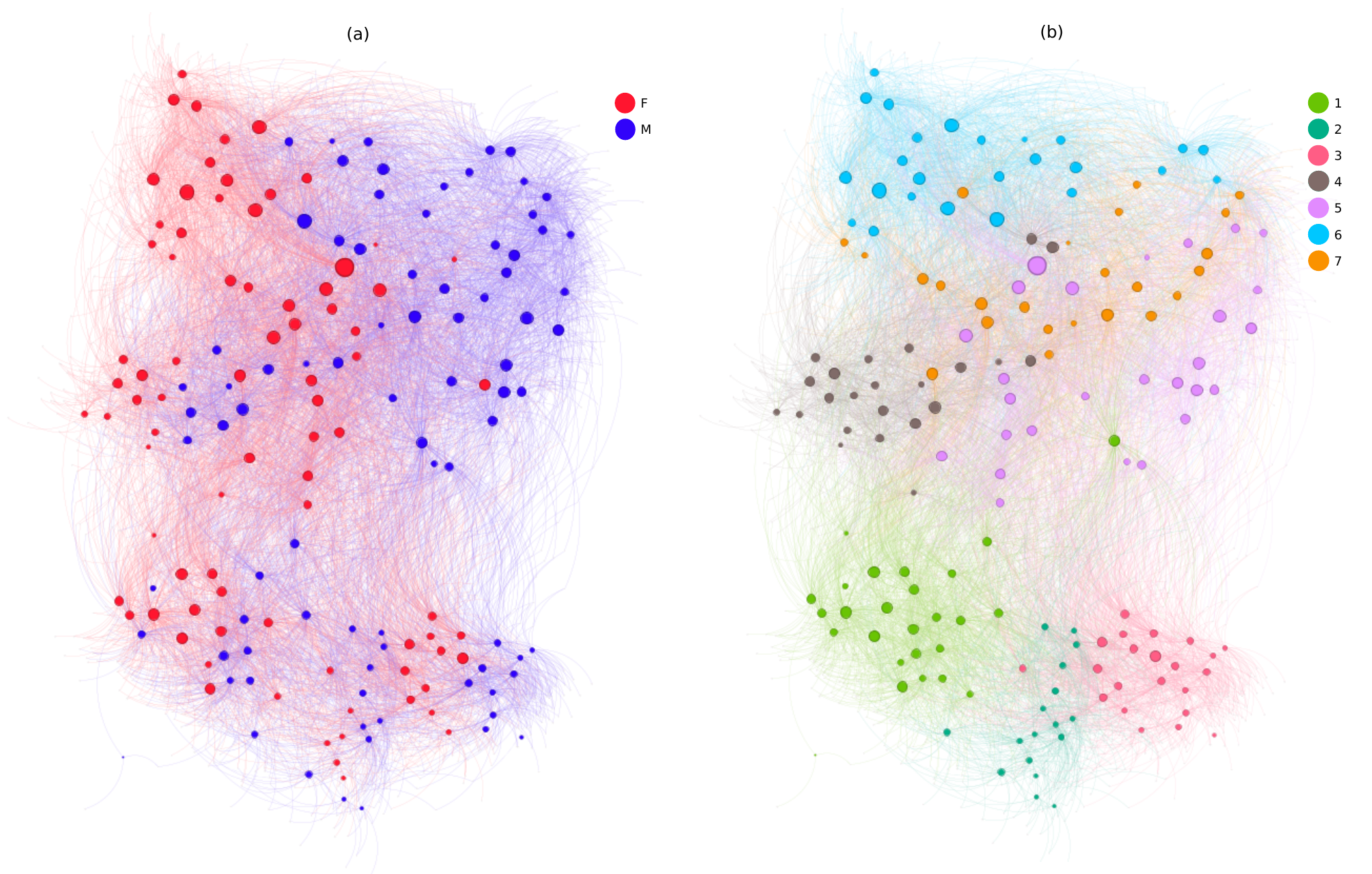}
		\caption{Visualizations of the temporal aggregated network during the out-of-class context. The size of the nodes corresponds to their degree. The layout is induced by the bipartite representation of interactions, where coloured nodes representing pupils are connected through ``phantom nodes'' representing group interactions (hyperedges). (a) Colors indicate the sex of pupils: blue for male and red for female. (b) Colors indicate class affiliation. The location of nodes remains the same between the two networks. The visualization has been produced using Gephi~\cite{bastian2009gephi}.}
		\label{fig:network_viz}
	\end{figure}
	
	Figure \ref{fig:network_viz} shows pictorial visualization of the resulting networks measured during the out-of-class context, aggregated over time. The positioning of the nodes is the same in both panels, but in Figure \ref{fig:network_viz}(a) nodes are colored according to the sex of the child, while in Figure \ref{fig:network_viz}(b) nodes are colored according to the class affiliation. We can see from this second case how, even during the out-of-class time, where in principle pupils are free to interact beyond their classmates, the network presents clusters: pupils from the same class preferentially interact with each other. Additionally, the network is polarized with respect to grade: lower grade classes are located at the bottom, while higher grade ones are located at the top. As already known~\cite{dai2022thesis}, Figure \ref{fig:network_viz}(a), reveals a clustering according to sex. Interestingly, different behaviors are observed depending on the age of the pupils. Children in classes 1, 2, 3, and 4 tend to cluster first by class, then within their classes by sex. Conversely, children in classes 5, 6, and 7 exhibit a different pattern, clustering first by sex and then by class. This qualitatively observed pattern suggests that younger children prioritize class-based groupings over gender, while older children exhibit stronger gender-based clustering, indicating a developmental shift in social dynamics. This change reflects previous work which finds that sex-based homophily increases with age in young children \cite{stehle_gender_2013, dai2022thesis}. Moreover, both younger and older children prefer to associate with others similar in age.
	
	\section{Results}
	\label{sec:results}
	In this section, we analyse the interplay of individual characteristics with behavioural social patterns starting from the simplest analysis at the level of individual nodes. We will then move to more complex dyadic and group measures that take into account the features of the nodes at different categorical levels. 
	
	The first and most fundamental quantity we consider is the total interaction time. Indeed, at the node level, the strength provides a simple and interpretable measure of the individual propensity to interact. Given the specific age group considered in this study, this represents a particularly interesting factor: it can be interpreted, in a dual manner, as an early indicator of both a tendency to interact and, conversely, as a proxy for a propensity to isolate.
	
	\subsection{Classification of interaction time using metadata}
	
	\begin{figure}
		\centering
		\includegraphics[width=\textwidth]{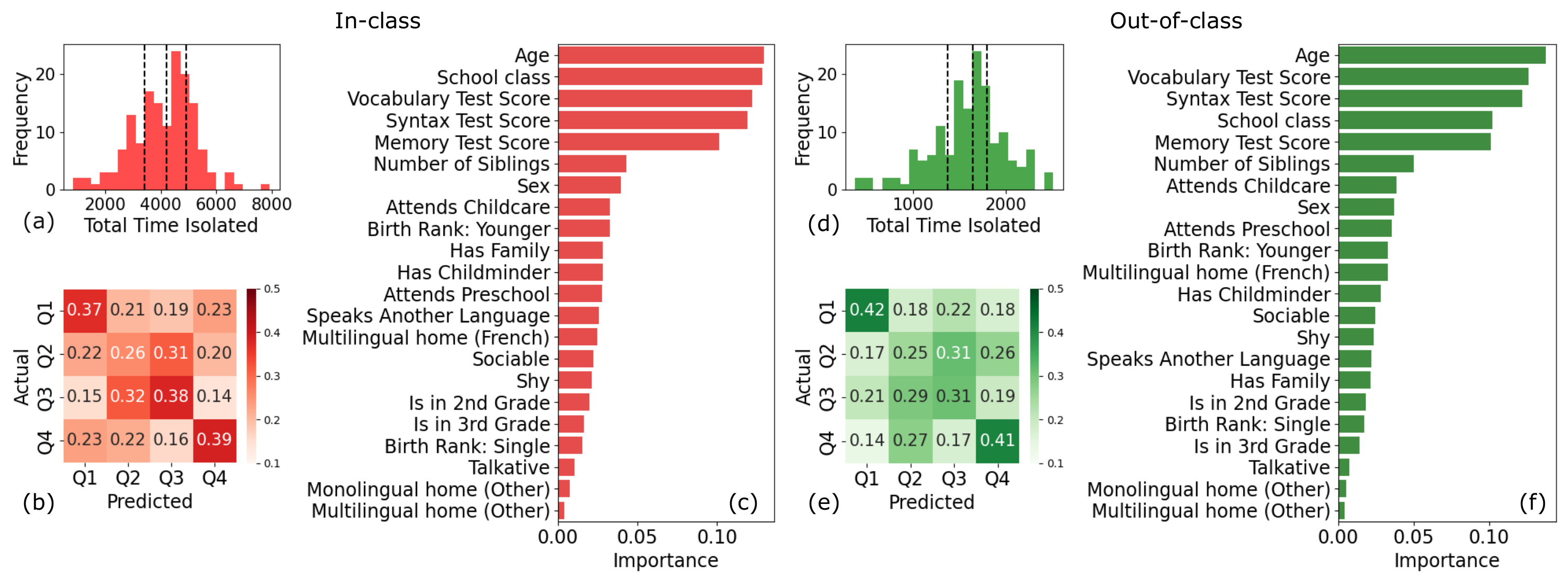}
		\caption{Classifying pupils into classes of social activity based on individual metadata during in-class (a-c) and out-of-class (d-f) time. (a,d) Pupils are divided into four classes of social activity based on individual time spent in social interactions. The 4 quantiles of the distributions are marked by dashed lines. (b, e) Confusion matrices associated to the classification task performed by a random forest model, that is able to distinguish the most and least socially active pupils. The bar plots (c,f) report the relevance the model assigns to each individual feature; the top 5 features are consistent across the two contexts.}
		\label{fig:classification_report}
	\end{figure}
	
	We split the distribution of interaction time into four quantile-based classes, emphasizing the first and last quantiles to represent the most and least socially active children, respectively [Figure ~\ref{fig:classification_report}(a,d)]. We then train a Random Forest classifier \citep{breiman2001random} on the metadata prepared as described in section ~\ref{sec:characteristics} to predict the quantile of each pupil. In order to improve the stability of our results, given the relatively small sample size, we employed stratified $k$-folding \citep{hastie2009elements} for training. The resulting confusion matrices, obtained by taking the average over 20 independent training realizations, are reported in Figure~\ref{fig:classification_report}(b) and (e) for the in-class and out-of-class settings, respectively. 
	
	Our analysis reveals that even though the metadata do not contain direct information on contacts, it is possible to use them to classify individuals according to their measured level of sociality. The signal, even though globally weak, enables good partial differentiation if one focuses on the two extreme quantiles ---which are also the most interesting ones as they depict the most and less active individuals. The signal is stronger in the out-of-class setting [Figure ~\ref{fig:classification_report}(e)], possibly due to the removal of spatial and logistic constraints imposed by the room and the teacher in the classrooms.
	We identify the most important features for classification, shown in Figure ~\ref{fig:classification_report}(c) and (f). Notice how the top five are consistent in the two settings: age, school-class and language test scores. This preliminary analysis suggests that individual characteristics that do not explicitly include levels of sociality are sufficient to detect the amount of interactions in preschool, in particular for the two most interesting subset of most and less active children.

	\subsection{Pairwise interaction analysis}
	
	We just saw that age can be a good predictor for duration of contact at the level of the single node. We would like to move a step further by taking into account the coupling at the level of links. Previous works have looked at the effect of age differences in classrooms on childrens' performance, but have not touched upon the social effects and differences associated with it~\cite{veenman1996effects, leroy2007revisiting}. 
	
\begin{figure}
		\centering
		\includegraphics[width=\textwidth]{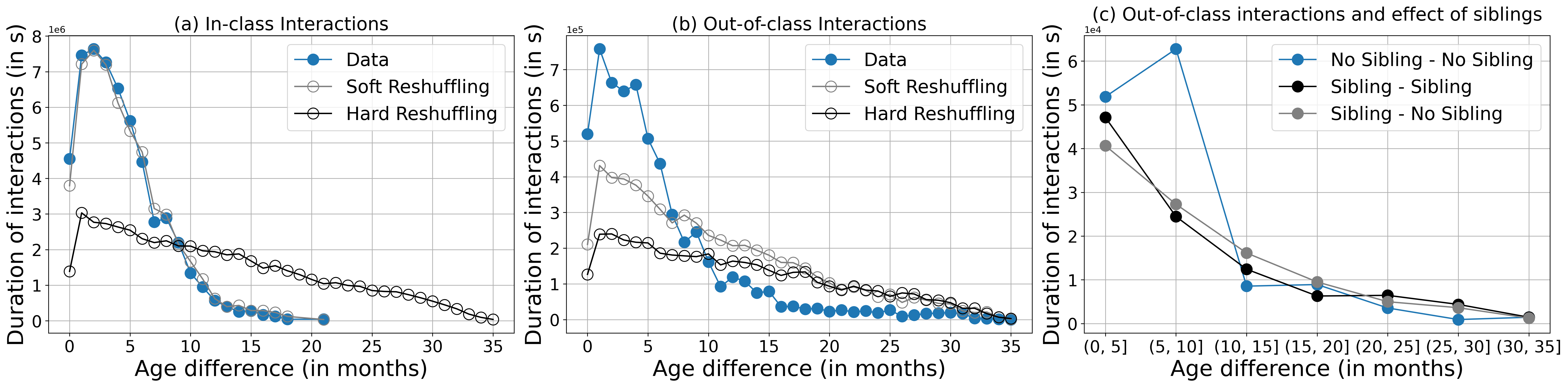}
		\caption{From the weighted interaction network of children aggregated across weeks, the duration of interactions is plotted against age difference of children interacting (a) during class time (b) during free time. The data is compared with 50 realizations each of hard and soft reshuffling models. The hard reshuffling does not preserve degree distribution nor the weight distribution while the soft reshuffling preserves degree distribution while shuffling the weights (duration of interaction) among edges. (c) Duration vs age difference for out-of-class interactions for pairs of nodes where both have siblings, both have no siblings, only one has siblings}
		\label{fig:interaction_duration_plot}
	\end{figure}

	To begin, we construct a weighted undirected network from all the in-class interactions happening between children of the same class, aggregated over all the 10 weeks of observation. In Figure~\ref{fig:interaction_duration_plot}(a) we plot the distribution of duration of in-class interactions against the age difference of the children involved in the considered pair. We see a peak (longer interactions) for pairs of pupils with less than half a year of age difference, but this quickly diminishes beyond that. To access the effect of network structure and/or the distribution of weights in the observed pattern, we use two reshuffling methods. The ``hard'' reshuffling removes all edges and reassigns them to random pairs of nodes (conserving only the total number of edges). This procedure completely destroys the network structure, and consequently also the duration distribution observed in the data. By contrast, the ``soft'' reshuffling removes all the weights (without changing the network structure) and reassigns them randomly to the edges. This method does not affect the patterns of interactions, indicating that the effect of how much you interact with somebody instead of another person in your network pales in comparison to the effect of network structure. We then repeat the same construction and reshuffling methods but for out-of-class interactions. While the distribution resembles that of in-class interactions, the extent to which you interact with people in your network matters significantly more compared to the null model. In Figure~\ref{fig:interaction_duration_plot}(b), we see that pupils with less than half a year age difference interact more than expected when compared to both the hard and the soft reshuffling null models. Interactions between children of larger age differences are also much less than expected from the soft reshuffling model.
	
	Having accessed these differences between in- and out-of-class interactions, we now want to look at possible differences between the level of sociality when exposed to different ``treatments''. From a social perspective, the most obvious change in mixed-grade classrooms is that students of different ages are forced, when in class, to interact with each other. When this restriction is removed, do these children interact differently?
	We can artificially explore the differences in ``removing'' this restriction by looking at interactions happening out-of-class but differentiating for children belonging to mixed- against single-grade classes. Notably, we do not observe significant differences in among the two. Even though a more rigorous check should be performed in further studies, this means that the distribution of duration of out-of-class contacts given an age difference, is not different if a pupil is exposed to larger age differences in class or not.
	
	We can look for alternative signals by leveraging information we have about the environment at home. For example, do children with siblings interact outside their age group more than only children? That is, do children who are exposed at home to other children with an age gap of roughly a year or more tend to interact more with varied age groups at school as well?
	In Figure~\ref{fig:interaction_duration_plot}(c), we plot the duration of out-of-class interactions against the age difference of interacting pairs of nodes for three cases: ({\it i}) both nodes have no siblings, ({\it ii}) both nodes have siblings, ({\it iii}) only one node has siblings. While this preliminary investigation seem to indicate that having a sibling does not seem to be associated with more interactions outside your age group, pupils who have no siblings seem to have a different pattern in the way duration depends on age difference: instead of observing a smooth decay with age difference, single children display a more pronounced peak for interactions with children within a year apart from each other, followed by an abrupt drop.

	\subsection{Higher-order interaction analysis}
	
	So far we have discussed, on the one hand, the interplay between the individual characteristics of a child and their individual propensity to interact, and on the other, the propensity of children of similar ages to spend more time together than those of larger age gaps. This latter analysis focused on the amount of time spent together by pairs of children. Nevertheless, as for many other systems, childrens' interactions can involve more than 2 pupils at the same time~\cite{lambiotte2019networks, battiston2020networks, torres2021and, bick2023higher}. Such group (non-dyadic) interactions have been observed in many social settings, such as offices, conferences, primary schools and high-schools, and it has been shown that the higher-order representations offered by hypergraphs, as opposed to pairwise graphs, can lead to interesting emerging phenomena~\cite{battiston2021physics}. Given the individual metadata available in this study, it is then natural to extend the study conducted so far and investigate how individual traits of group members influence the overall time spent together in a given group gathering --beyond the pairs.
	
	To address this question, we deduced the group interactions among children from the proximity data discussed above. Interaction data are recorded in the form of tuples $((i,j),t)$ between pairs of children occurring at a specific time $t$, where $(i,j)$ is the pair of nodes (pupils) interacting and $t$ is the time at which the interaction was observed. Given the temporal granularity, it is however possible to naively construct the group interactions by looking at the cliques formed by pairwise interactions occurring at the same timestamp, (following the procedure of Refs.~\cite{iacopini2019simplicial, cencetti2021temporal}). For example, if at a specific timestamp $t$ we observe that in the contact data nodes $i$, $j$ and $k$ are all mutually connected by a link, i.e. we observe $((i,j),t)$, $((j,k),t)$ and $((i,k),t)$, then the three nodes are considered as jointly interacting in a group at time $t$. The so-obtained network can then be considered what has been recently addressed as a higher-order network~\cite{battiston2020networks}, where group interactions among children are represented as hyperlinks, that are links that connect an arbitrary number of nodes. Similarly to the pairwise case, we then assign to each hyperlink a weight which accounts for the total temporal duration the hyperlink was observed in the contact data. Hyperlinks with null weights (i.e., never observed) were not considered in this analysis. 
	
	We integrate individual traits belonging to different group member into a measure of affinity for the whole group. In particular, we considered two individual characteristics that were included in the previous analyses, i.e. age and results in development skill tests. While age is a simple measure given in months, test scores require some pre-processing. We first normalise the test results such that the score obtained by children is always between 0 and 1. Then, for each pupil, we store these results into a 3-dimensional vector, where each entry corresponds to the result obtained in the test on vocabulary, syntax and memory.
	We can now quantify {\it group affinity} in terms of how further apart are its members with respect to the considered quantity. A natural way of doing this is to consider individual distances from the barycenter of the group. We thus measure the diversity in age and development skills test scores by calculating the sum of the absolute distances of each node from the barycentric coordinate, and then dividing this sum by the number of group members. As per previous analyses, we split interactions into in-class and out-of-class, to distinguish between the two different social contexts in which pupils interact.
	
\begin{figure}[t]
		\centering
		\includegraphics[width=0.8\textwidth]{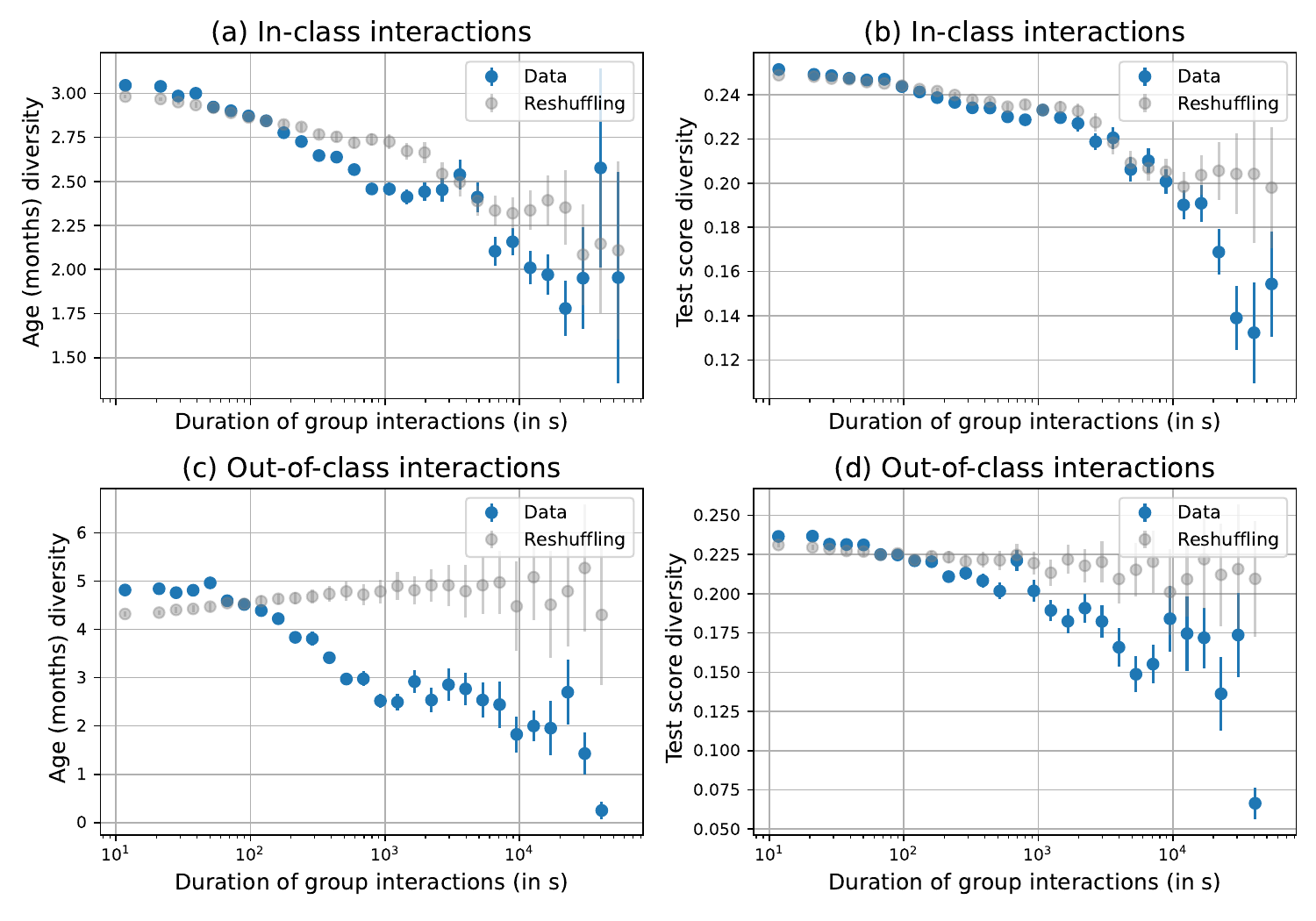}
		\caption{The duration of group interactions observed in pupils interacting during class (a,b) and out of class (c,d) is plotted against group diversity measured from individual characteristics: age (a,c) and language development test scores (b,d). Results obtained from empirical data (blue) are compared to those of randomized null model (gray), where alter group members are taken at random from a different group of the same size over 10 different realizations.}
		\label{fig:group_interaction_duration_plot}
	\end{figure}

	In Figure~\ref{fig:group_interaction_duration_plot} we plot group diversity as a function of the group duration for the two different individual traits and the two different contexts of interaction considered. We observe that longer-lasting groups tend to be less distant, thus group affinity is higher, both in terms of age and development skill performances. Results obtained from empirical data are compared with randomized reference models, where the distance between a node and its group is computed instead using different nodes taken at random from another group of the same size. 
	In the in-class context [Figure~\ref{fig:group_interaction_duration_plot}(a-b)], we see that the empirical groups that last for long present lower diversity that those from the reshuffled model. This trend is much more evident in the case of age diversity (a) than in the case of development skill tests (b), where appreciable differences only appear for the longer-lasting groups. 
	Signals are stronger in the out-of-class context [Figure~\ref{fig:group_interaction_duration_plot}(c-d)], where we can see clear differences of trends between empirical data and the reshuffled cases. The group diversity after reshuffling remains constant as group duration increases, in striking contrast to the decreasing trend observed in real-world data.
	We conclude that, whenever pupils are left free to interact (out-of-class), they gather in groups whose duration depends on group affinity in terms of individual age and development skill test performances, independently from their size.
	
	\subsection{Node attributes and network measures}
	
	\begin{figure}
		\centering
		\includegraphics[width=0.9\textwidth]{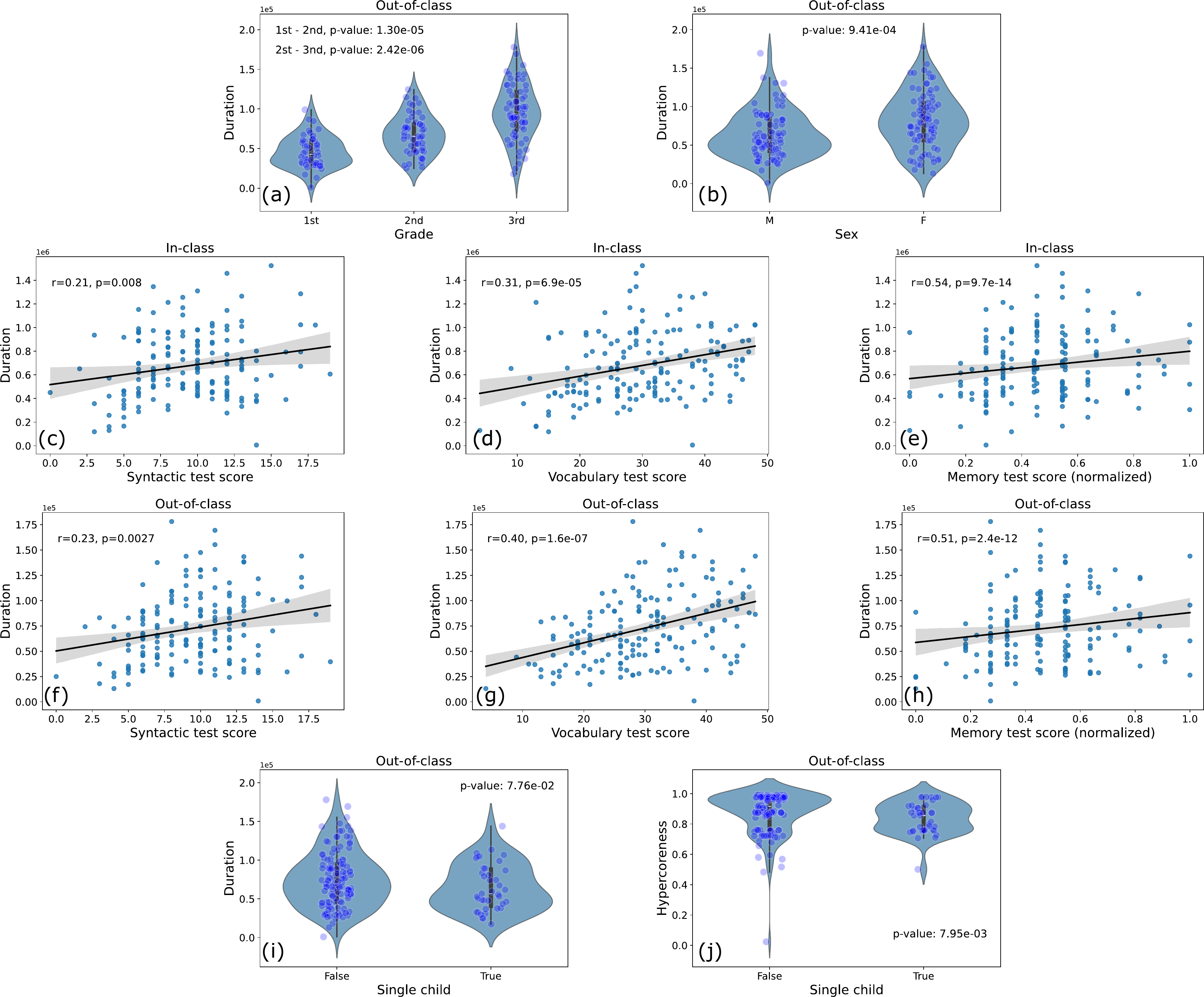}
		\caption{Distribution of contact duration across for children belonging to different grades (a) and sex categories (b). (c-h) Association between duration of contacts and language development test scores in different contexts. Distribution of contact duration (i) and hypercoreness centrality (j) for children with and without siblings.}
		\label{fig:meta-network-clara}
	\end{figure}
	
	In this final section, we consider node-level measures of interaction computed for individual pupils and look for differences in their distributions when accounting for different categories of individual traits. In particular, we focus on the same characteristics used in the previous sections: sex, grade, development skill test scores, and number of siblings.
	
	We find that the overall the interaction time of female in out-of-class settings is higher than the one of male children. This is shown in Figure~\ref{fig:meta-network-clara}(b) where the two strength distributions are compared across the two sex categories (significance was checked using a 1-sided Mann–Whitney U test, $p<0.01$).
	We also notice that total duration of out-of-class interactions increases with age. This is displayed in Figure~\ref{fig:meta-network-clara}(a) where the strength distribution for both female and male is plotted for different grade levels. Even if this is not shown in the figure, a temporal analysis shows that this effect seems to be more pronounced in the first quarter in the academic year. This could be due to the differences in the acquaintance network prior to the beginning of the data collection (stronger for 1st graders). 
	
	We then turn our attention to language development measures in association with duration of contacts in both in-class and out-of-class settings. We first discard the results of anchor questions ---since they are positively correlated with grade, which we have already shown to be correlated with strength. In all settings we find [Figure~\ref{fig:meta-network-clara}(c-h)] a significant positive correlation for all test scores, in particular for memory and vocabulary tests (which are not correlated with one another).
	
	Finally, we focus on differences in out-of-class contacts for children with and without siblings both in terms of duration of interactions and their patterns (how interactions are arranged across groups). In Figure~\ref{fig:meta-network-clara}(i) we compare distributions of contact duration across the two categories, finding no significant differences across the two. However, going beyond simple contact duration, differences emerge. This is shown in Figure~\ref{fig:meta-network-clara}(j), where we compare the hypercoreness values of the nodes. Hypercoreness is a recently-developed measure of centrality for higher-order networks \cite{mancastroppa2023hyper, mancastroppa2024structural} that quantifies the extent of nodes' interactions within groups, considering at the same time number of different groups and group sizes. High hypercoreness values correspond to nodes that interact within many large-sized groups that contain high-degree nodes~\footnote{More formally, we start by performing a $(k, m)$-hypercore decomposition \cite{liu2020efficient, mancastroppa2023hyper}. The $(k, m)$-hypercore decomposition is analogous to a $k$-core decomposition for graphs, involving the recursive removal of nodes with degree $k_i < k$ and groups (or hyper-edges) with size $m_e < m$. The resulting $(k,m)$-hypercore is the maximal connected sub-network where all nodes belong to at least $k$ distinct groups and all groups of size at least $m$. The hypercoreness centrality  of a given node $i$ is defined as:
		$$ R(i) = \sum_{m=2}^{M} \frac{ C_m(i) }{N_m k_{max}^{m}}$$
		where the $m$-core number, denoted as $C_m(i)$, is the value $k$ such that $i$ belongs to the $(k,m)$-hypercore but not the $(k+1,m)$-hypercore, $k_{max}^{m}$ is the maximum value of $k$ such that the $(k,m)$-hypercore is not empty and $N_m$ is the number of groups of size $m$ in the original network.}.

	\section{Discussion}
	\label{sec:discussion}
	
	We investigated the interplay of individual characteristics of children at preschool with their longitudinal patterns of face-to-face interactions automatically collected across different classes and contexts of interactions. Starting from the amount of sociality as given by the total time spent interacting/in isolation, we showed it is possible to make predictions about the social activity of children using individual characteristics. 
	
	We subsequently used random forest methods to identify the key individual traits playing a role in interactions between students in and out of class as well as across mixed-grade classes. The main traits found were age, sex, number of siblings and test scores (vocabulary, syntax and memory scores). We identified an increasing preference with age for children to interact with same-sex others as well as a preference for children to associate with classmates outside of class. Furthermore, children were found to prefer to associate with others similar in age to themselves. 
	
	Our investigation into the age-based in-class and out-of class interaction patterns among children reveals that the age difference significantly influences the duration of in-class interactions, with the highest interactions occurring among children with less than half a year age difference. To understand the role of network structure and weight distribution in these patterns, we employed two reshuffling methods: hard reshuffling (randomly reassigning edges) and soft reshuffling (randomly reassigning weights while preserving network structure). The hard reshuffling did not replicate the observed duration distribution, indicating the importance of network structure, whereas the soft reshuffling showed that interaction duration within the network was less critical. Out-of-class interactions mirrored the in-class patterns but highlighted a significantly greater importance of interaction duration even when the network structure is preserved. Further, we examined whether having siblings affected childrens' interactions across age groups. The analysis indicated no significant difference in interaction patterns based on sibling presence, except for children without siblings who showed longer interactions within their age group. These preliminary results suggest that exposure to mixed-age interactions within the family does not significantly influence the age gap in social interactions at school ---apart for a different decreasing trend on how contact duration decreases as the gap increases. Nevertheless, additional comprehensive causal work is necessary to make this inference.
	
	At the group level, we found differences across children with or without siblings in terms of node centrality in the higher-order networks. In fact, even though no significant differences emerge when comparing group durations, children with siblings display higher hypercoreness~\cite{mancastroppa2023hyper}, meaning that they engage in more groups of larger sizes. Along this line, an interesting directions to investigate would focus on temporal hypercoreness~\cite{mancastroppa2024structural}, leveraging the longitudinal nature of the dataset, or more complex local interactions patterns as hypermotifs~\cite{lotito2022higher}.
	
	Despite its prevalence, the French double-grade class system has been criticised and linked to poor performance outcomes due to teachers needing to switch attention between two teaching groups. However some benefits of this system presented were tutoring, imitation and joint supervision. Future work could explore how age differences, number of students and teaching abilities lead to positive or negative interactions and outcomes~\cite{suchaut2010efficacite}. In the context of early cognitive development, an important direction of future work could couple neuro-cognitive measures with measures of sociability to learn more about how brain development changes the nature of social interactions.
	
	Another natural direction to explore in future work is the use of affinity measures for individual characteristics within group interactions based on the higher-order definition recently presented in Ref.~\cite{veldt2023combinatorial}, but generalised to account for labels that can take more than two values. More generally, our preliminary study calls for a more comprehensive and exhaustive investigation of homophilic~\cite{altenburger2018monophily} ---and monophilic~\cite{altenburger2018monophily}--- patterns of group formation and evolution. These signals could then be used to inform mechanistic model of higher-order social networks~\cite{petri2018simplicial, gallo2024higher, iacopini2023temporal, gallo2024higher_f2f}. In fact, recent studies have found consistent dynamical patterns of individual group transitions, group formation and disaggregation phenomena in both preschool and university settings during different activity types (in-class, out-of-class, and weekend)~\cite{iacopini2023temporal}. The observed phenomena could be replicated by a synthetic model describing the dynamics of individuals forming groups of different sizes and navigating through them, using a mechanism of short-term memory for group duration (“long gets longer” effect) and long-term memory for social contacts. Going beyond these simple signature of recurrent social contact, the individual preferences analysed in this study could be used to complement and further improve these mechanistic dynamics of social interactions. 
	
	\section{Acknowledgements}
	This work is the output of the Complexity72h workshop, held at the Universidad Carlos III de Madrid in Leganés, Spain, 24-28 June 2024, \url{https://www.complexity72h.com}. We thank Alain Barrat for the insightful discussions and contributions during the workshop. The higher-order interaction analysis was performed using the XGI library~\cite{landry2023xgi}.

    \bibliographystyle{unsrtnat}

\begin{thebibliography}{72}
		\providecommand{\natexlab}[1]{#1}
		\providecommand{\url}[1]{\texttt{#1}}
		\expandafter\ifx\csname urlstyle\endcsname\relax
		\providecommand{\doi}[1]{doi: #1}\else
		\providecommand{\doi}{doi: \begingroup \urlstyle{rm}\Url}\fi
		
		\bibitem[Albert and Barab{\'a}si(2002)]{albert2002statistical}
		R{\'e}ka Albert and Albert-L{\'a}szl{\'o} Barab{\'a}si.
		\newblock Statistical mechanics of complex networks.
		\newblock \emph{Rev. Mod. Phys.}, 74\penalty0 (1):\penalty0 47, 2002.
		\newblock \doi{https://doi.org/10.1103/RevModPhys.74.47}.
		
		\bibitem[Newman(2003)]{newman2003structure}
		Mark~EJ Newman.
		\newblock The structure and function of complex networks.
		\newblock \emph{SIAM review}, 45\penalty0 (2):\penalty0 167--256, 2003.
		\newblock \doi{https://doi.org/10.1137/S003614450342480}.
		
		\bibitem[Latora et~al.(2017)Latora, Nicosia, and
		Russo]{latora_nicosia_russo_2017}
		V.~Latora, V.~Nicosia, and G.~Russo.
		\newblock \emph{Complex Networks: Principles, Methods and Applications}.
		\newblock Complex Networks: Principles, Methods and Applications. Cambridge
		University Press, 2017.
		\newblock ISBN 9781107103184.
		\newblock URL \url{https://books.google.it/books?id=qV0yDwAAQBAJ}.
		
		\bibitem[Barrat et~al.(2008)Barrat, Barth{\'e}lemy, and
		Vespignani]{barrat2008dynamical}
		A.~Barrat, M.~Barth{\'e}lemy, and A.~Vespignani.
		\newblock \emph{Dynamical Processes on Complex Networks}.
		\newblock Cambridge University Press, 2008.
		\newblock ISBN 9780521879507.
		\newblock URL \url{https://books.google.at/books?id=TmgePn9uQD4C}.
		
		\bibitem[Vespignani(2012)]{vespignani2012modelling}
		Alessandro Vespignani.
		\newblock Modelling dynamical processes in complex socio-technical systems.
		\newblock \emph{Nature Physics}, 8\penalty0 (1):\penalty0 32--39, 2012.
		
		\bibitem[Euler(1741)]{euler1741solutio}
		Leonhard Euler.
		\newblock Solutio problematis ad geometriam situs pertinentis.
		\newblock \emph{Commentarii academiae scientiarum Petropolitanae}, pages
		128--140, 1741.
		
		\bibitem[Wasserman and Faust(1994)]{wasserman1994social}
		Stanley Wasserman and Katherine Faust.
		\newblock \emph{Social Network Analysis : {{Methods}} and Applications
			(Structural Analysis in the Social Sciences)}.
		\newblock {Cambridge University Press}, 1994.
		\newblock ISBN 0-521-38707-8.
		
		\bibitem[Castellano et~al.(2009)Castellano, Fortunato, and
		Loreto]{castellano2009statistical}
		Claudio Castellano, Santo Fortunato, and Vittorio Loreto.
		\newblock Statistical physics of social dynamics.
		\newblock \emph{Rev. Mod. Phys.}, 81\penalty0 (2):\penalty0 591, 2009.
		\newblock \doi{https://doi.org/10.1103/RevModPhys.81.591}.
		
		\bibitem[Baronchelli(2018)]{baronchelli2018emergence}
		Andrea Baronchelli.
		\newblock The emergence of consensus: A primer.
		\newblock \emph{R. Soc Open Sci}, 5\penalty0 (2):\penalty0 172189, 2018.
		
		\bibitem[Axelrod and Axelrod(1984)]{axelrod1984evolution}
		R.~Axelrod and R.M. Axelrod.
		\newblock \emph{The Evolution of Cooperation}.
		\newblock Basic Books. {Basic Books}, 1984.
		\newblock ISBN 978-0-465-02121-5.
		
		\bibitem[Mitchell(1973)]{mitchell1973networks}
		J~Clyde Mitchell.
		\newblock \emph{Networks, norms and institutions}.
		\newblock Mouton., 1973.
		
		\bibitem[Centola and Macy(2007)]{centola2007complex}
		Damon Centola and Michael Macy.
		\newblock Complex contagions and the weakness of long ties.
		\newblock \emph{American journal of Sociology}, 113\penalty0 (3):\penalty0
		702--734, 2007.
		
		\bibitem[Nowak et~al.(1990)Nowak, Szamrej, and Latan{\'e}]{nowak1990private}
		Andrzej Nowak, Jacek Szamrej, and Bibb Latan{\'e}.
		\newblock From private attitude to public opinion: A dynamic theory of social
		impact.
		\newblock \emph{Psychol. Rev.}, 97\penalty0 (3):\penalty0 362, 1990.
		
		\bibitem[Sznajd-Weron and Sznajd(2000)]{sznajd2000opinion}
		Katarzyna Sznajd-Weron and Jozef Sznajd.
		\newblock Opinion evolution in closed community.
		\newblock \emph{International Journal of Modern Physics C}, 11\penalty0
		(06):\penalty0 1157--1165, 2000.
		
		\bibitem[Isella et~al.(2011)Isella, Stehl{\'e}, Barrat, Cattuto, Pinton, and
		Van~den Broeck]{isella2011s}
		Lorenzo Isella, Juliette Stehl{\'e}, Alain Barrat, Ciro Cattuto,
		Jean-Fran{\c{c}}ois Pinton, and Wouter Van~den Broeck.
		\newblock What's in a crowd? analysis of face-to-face behavioral networks.
		\newblock \emph{Journal of theoretical biology}, 271\penalty0 (1):\penalty0
		166--180, 2011.
		
		\bibitem[Barrat et~al.(2014)Barrat, Cattuto, Tozzi, Vanhems, and
		Voirin]{barrat2014measuring}
		Alain Barrat, Ciro Cattuto, Alberto~Eugenio Tozzi, Philippe Vanhems, and
		Nicolas Voirin.
		\newblock Measuring contact patterns with wearable sensors: methods, data
		characteristics and applications to data-driven simulations of infectious
		diseases.
		\newblock \emph{Clin. Microbiol. Infect.}, 20\penalty0 (1):\penalty0 10--16,
		2014.
		\newblock \doi{https://doi.org/10.1111/1469-0691.12472}.
		
		\bibitem[Mastrandrea et~al.(2015{\natexlab{a}})Mastrandrea, Fournet, and
		Barrat]{mastrandrea2015contact}
		Rossana Mastrandrea, Julie Fournet, and Alain Barrat.
		\newblock Contact patterns in a high school: a comparison between data
		collected using wearable sensors, contact diaries and friendship surveys.
		\newblock \emph{PloS one}, 10\penalty0 (9):\penalty0 e0136497,
		2015{\natexlab{a}}.
		
		\bibitem[Cristani et~al.(2011)Cristani, Paggetti, Vinciarelli, Bazzani,
		Menegaz, and Murino]{cristani2011towards}
		Marco Cristani, Giulia Paggetti, Alessandro Vinciarelli, Loris Bazzani, Gloria
		Menegaz, and Vittorio Murino.
		\newblock Towards computational proxemics: Inferring social relations from
		interpersonal distances.
		\newblock In \emph{2011 IEEE Third International Conference on Privacy,
			Security, Risk and Trust and 2011 IEEE Third International Conference on
			Social Computing}, pages 290--297. IEEE, 2011.
		
		\bibitem[Holme and Saram{\"a}ki(2012)]{holme2012temporal}
		Petter Holme and Jari Saram{\"a}ki.
		\newblock Temporal networks.
		\newblock \emph{Phys. Rep.}, 519\penalty0 (3):\penalty0 97--125, 2012.
		\newblock \doi{https://doi.org/10.1016/j.physrep.2012.03.001}.
		
		\bibitem[G{\'e}nois et~al.(2019)G{\'e}nois, Zens, Lechner, Rammstedt, and
		Strohmaier]{genois2019building}
		Mathieu G{\'e}nois, Maria Zens, Clemens Lechner, Beatrice Rammstedt, and Markus
		Strohmaier.
		\newblock Building connections: How scientists meet each other during a
		conference.
		\newblock \emph{arXiv preprint arXiv:1901.01182}, 2019.
		
		\bibitem[Battiston et~al.(2020)Battiston, Cencetti, Iacopini, Latora, Lucas,
		Patania, Young, and Petri]{battiston2020networks}
		F.~Battiston, G.~Cencetti, I.~Iacopini, V.~Latora, M.~Lucas, A.~Patania, J.-G.
		Young, and G.~Petri.
		\newblock Networks beyond pairwise interactions: {{Structure}} and dynamics.
		\newblock \emph{Phys. Rep.}, 874:\penalty0 1--92, 2020.
		\newblock \doi{10.1016/j.physrep.2020.05.004}.
		
		\bibitem[Battiston et~al.(2021)Battiston, Amico, Barrat, Bianconi, Ferraz~de
		Arruda, Franceschiello, Iacopini, K{\'e}fi, Latora, Moreno, Murray, Peixoto,
		Vaccarino, and Petri]{battiston2021physics}
		F.~Battiston, E.~Amico, A.~Barrat, G.~Bianconi, G.~Ferraz~de Arruda,
		B.~Franceschiello, I.~Iacopini, S.~K{\'e}fi, V.~Latora, Y.~Moreno, M.~Murray,
		T.~Peixoto, F.~Vaccarino, and G.~Petri.
		\newblock The physics of higher-order interactions in complex systems.
		\newblock \emph{Nat. Phys.}, 17\penalty0 (10):\penalty0 1093--1098, 2021.
		\newblock \doi{10.1038/s41567-021-01371-4}.
		
		\bibitem[Torres et~al.(2021)Torres, Blevins, Bassett, and
		Eliassi-Rad]{torres2021and}
		Leo Torres, Ann~S Blevins, Danielle Bassett, and Tina Eliassi-Rad.
		\newblock The why, how, and when of representations for complex systems.
		\newblock \emph{SIAM Rev.}, 63\penalty0 (3):\penalty0 435--485, 2021.
		\newblock \doi{https://doi.org/10.1137/20M1355896}.
		
		\bibitem[Bick et~al.(2023)Bick, Gross, Harrington, and Schaub]{bick2023higher}
		Christian Bick, Elizabeth Gross, Heather~A Harrington, and Michael~T Schaub.
		\newblock What are higher-order networks?
		\newblock \emph{SIAM Rev.}, 65\penalty0 (3):\penalty0 686--731, 2023.
		\newblock \doi{https://doi.org/10.1137/21M1414024}.
		
		\bibitem[Stehl{\'e} et~al.(2010)Stehl{\'e}, Barrat, and
		Bianconi]{stehle2010dynamical}
		Juliette Stehl{\'e}, Alain Barrat, and Ginestra Bianconi.
		\newblock Dynamical and bursty interactions in social networks.
		\newblock \emph{Phys. Rev. E}, 81\penalty0 (3):\penalty0 035101, 2010.
		\newblock \doi{10.1103/PhysRevE.81.035101}.
		
		\bibitem[Zhao et~al.(2011)Zhao, Stehl{\'e}, Bianconi, and
		Barrat]{zhao2011social}
		Kun Zhao, Juliette Stehl{\'e}, Ginestra Bianconi, and Alain Barrat.
		\newblock Social network dynamics of face-to-face interactions.
		\newblock \emph{Phys. Rev. E}, 83\penalty0 (5):\penalty0 056109, 2011.
		\newblock \doi{10.1103/PhysRevE.83.056109}.
		
		\bibitem[Perra et~al.(2012)Perra, Gon{\c{c}}alves, Pastor-Satorras, and
		Vespignani]{perra2012activity}
		Nicola Perra, Bruno Gon{\c{c}}alves, Romualdo Pastor-Satorras, and Alessandro
		Vespignani.
		\newblock Activity driven modeling of time varying networks.
		\newblock \emph{Scientific reports}, 2\penalty0 (1):\penalty0 469, 2012.
		
		\bibitem[Starnini et~al.(2013)Starnini, Baronchelli, and
		Pastor-Satorras]{starnini2013modeling}
		Michele Starnini, Andrea Baronchelli, and Romualdo Pastor-Satorras.
		\newblock Modeling human dynamics of face-to-face interaction networks.
		\newblock \emph{Phys. Rev. Lett.}, 110\penalty0 (16):\penalty0 168701, 2013.
		\newblock \doi{10.1103/PhysRevLett.110.168701}.
		
		\bibitem[Vestergaard et~al.(2014)Vestergaard, G{\'e}nois, and
		Barrat]{vestergaard2014memory}
		Christian~L Vestergaard, Mathieu G{\'e}nois, and Alain Barrat.
		\newblock How memory generates heterogeneous dynamics in temporal networks.
		\newblock \emph{Phys. Rev. E}, 90\penalty0 (4):\penalty0 042805, 2014.
		\newblock \doi{https://doi.org/10.1103/PhysRevE.90.042805}.
		
		\bibitem[Karsai et~al.(2014)Karsai, Perra, and Vespignani]{karsai2014time}
		M{\'a}rton Karsai, Nicola Perra, and Alessandro Vespignani.
		\newblock Time varying networks and the weakness of strong ties.
		\newblock \emph{Sci. Rep.}, 4\penalty0 (1):\penalty0 1--7, 2014.
		\newblock \doi{https://doi.org/10.1038/srep04001}.
		
		\bibitem[Nadini et~al.(2018)Nadini, Sun, Ubaldi, Starnini, Rizzo, and
		Perra]{nadini2018epidemic}
		Matthieu Nadini, Kaiyuan Sun, Enrico Ubaldi, Michele Starnini, Alessandro
		Rizzo, and Nicola Perra.
		\newblock Epidemic spreading in modular time-varying networks.
		\newblock \emph{Sci. Rep.}, 8\penalty0 (1):\penalty0 1--11, 2018.
		\newblock \doi{https://doi.org/10.1038/s41598-018-20908-x}.
		
		\bibitem[Le~Bail et~al.(2023)Le~Bail, G\'enois, and
		Barrat]{lebail2023modelling}
		Didier Le~Bail, Mathieu G\'enois, and Alain Barrat.
		\newblock Modeling framework unifying contact and social networks.
		\newblock \emph{Phys. Rev. E}, 107:\penalty0 024301, Feb 2023.
		\newblock \doi{10.1103/PhysRevE.107.024301}.
		\newblock URL \url{https://link.aps.org/doi/10.1103/PhysRevE.107.024301}.
		
		\bibitem[Petri and Barrat(2018)]{petri2018simplicial}
		Giovanni Petri and Alain Barrat.
		\newblock Simplicial activity driven model.
		\newblock \emph{Phys. Rev. Lett.}, 121\penalty0 (22):\penalty0 228301, 2018.
		\newblock \doi{10.1103/PhysRevLett.121.228301}.
		
		\bibitem[Gallo et~al.(2024{\natexlab{a}})Gallo, Lacasa, Latora, and
		Battiston]{gallo2024higher}
		Luca Gallo, Lucas Lacasa, Vito Latora, and Federico Battiston.
		\newblock Higher-order correlations reveal complex memory in temporal
		hypergraphs.
		\newblock \emph{Nature Communications}, 15\penalty0 (1):\penalty0 4754,
		2024{\natexlab{a}}.
		
		\bibitem[Iacopini et~al.(2023)Iacopini, Karsai, and
		Barrat]{iacopini2023temporal}
		Iacopo Iacopini, M{\'a}rton Karsai, and Alain Barrat.
		\newblock The temporal dynamics of group interactions in higher-order social
		networks.
		\newblock \emph{arXiv preprint arXiv:2306.09967}, 2023.
		
		\bibitem[Kontro and G{\'e}nois(2020)]{kontro2020combining}
		Inkeri Kontro and Mathieu G{\'e}nois.
		\newblock Combining surveys and sensors to explore student behaviour.
		\newblock \emph{Education Sciences}, 10\penalty0 (3):\penalty0 68, 2020.
		
		\bibitem[Dai et~al.(2022)Dai, Bouchet, Karsai, Chevrot, Fleury, and
		Nardy]{dai2022longitudinal}
		Sicheng Dai, H{\'e}l{\`e}ne Bouchet, M{\'a}rton Karsai, Jean-Pierre Chevrot,
		Eric Fleury, and Aur{\'e}lie Nardy.
		\newblock Longitudinal data collection to follow social network and language
		development dynamics at preschool.
		\newblock \emph{Scientific Data}, 9\penalty0 (1):\penalty0 777, 2022.
		
		\bibitem[G{\'e}nois et~al.(2023)G{\'e}nois, Zens, Oliveira, Lechner, Schaible,
		and Strohmaier]{genois2023combining}
		Mathieu G{\'e}nois, Maria Zens, Marcos Oliveira, Clemens~M Lechner, Johann
		Schaible, and Markus Strohmaier.
		\newblock Combining sensors and surveys to study social interactions: A case of
		four science conferences.
		\newblock \emph{Pers. Sci.}, 4:\penalty0 1--24, 2023.
		\newblock \doi{https://doi.org/10.5964/ps.9957}.
		
		\bibitem[Stehl{\'e} et~al.(2013)Stehl{\'e}, Charbonnier, Picard, Cattuto, and
		Barrat]{stehle_gender_2013}
		Juliette Stehl{\'e}, Fran{\c c}ois Charbonnier, Tristan Picard, Ciro Cattuto,
		and Alain Barrat.
		\newblock Gender homophily from spatial behavior in a primary school: {A}
		sociometric study.
		\newblock \emph{Social Networks}, 35\penalty0 (4):\penalty0 604--613, October
		2013.
		\newblock ISSN 03788733.
		\newblock \doi{10.1016/j.socnet.2013.08.003}.
		\newblock URL
		\url{https://linkinghub.elsevier.com/retrieve/pii/S0378873313000737}.
		
		\bibitem[Mastrandrea et~al.(2015{\natexlab{b}})Mastrandrea, Fournet, and
		Barrat]{mastrandrea_contact_2015}
		Rossana Mastrandrea, Julie Fournet, and Alain Barrat.
		\newblock Contact {Patterns} in a {High} {School}: {A} {Comparison} between
		{Data} {Collected} {Using} {Wearable} {Sensors}, {Contact} {Diaries} and
		{Friendship} {Surveys}.
		\newblock \emph{PLOS ONE}, 10\penalty0 (9):\penalty0 e0136497, September
		2015{\natexlab{b}}.
		\newblock ISSN 1932-6203.
		\newblock \doi{10.1371/journal.pone.0136497}.
		\newblock URL \url{https://dx.plos.org/10.1371/journal.pone.0136497}.
		
		\bibitem[McClelland et~al.(2007)McClelland, Cameron, Connor, Farris, Jewkes,
		and Morrison]{mcclelland2007links}
		Megan~M McClelland, Claire~E Cameron, Carol~McDonald Connor, Carrie~L Farris,
		Abigail~M Jewkes, and Frederick~J Morrison.
		\newblock Links between behavioral regulation and preschoolers' literacy,
		vocabulary, and math skills.
		\newblock \emph{Developmental psychology}, 43\penalty0 (4):\penalty0 947, 2007.
		
		\bibitem[Denham(2006)]{denham2006social}
		Susanne~A Denham.
		\newblock Social-emotional competence as support for school readiness: What is
		it and how do we assess it?
		\newblock \emph{Early education and development}, 17\penalty0 (1):\penalty0
		57--89, 2006.
		
		\bibitem[Birch and Ladd(1997)]{birch1997teacher}
		Sondra~H Birch and Gary~W Ladd.
		\newblock The teacher-child relationship and children's early school
		adjustment.
		\newblock \emph{Journal of school psychology}, 35\penalty0 (1):\penalty0
		61--79, 1997.
		
		\bibitem[McPherson et~al.(2001)McPherson, Smith-Lovin, and
		Cook]{mcpherson2001birds}
		Miller McPherson, Lynn Smith-Lovin, and James~M Cook.
		\newblock Birds of a feather: Homophily in social networks.
		\newblock \emph{Annual review of sociology}, 27\penalty0 (1):\penalty0
		415--444, 2001.
		
		\bibitem[Mayhew et~al.(1995)Mayhew, McPherson, Rotolo, and
		Smith-Lovin]{mayhew1995sex}
		Bruce~H Mayhew, J~Miller McPherson, Thomas Rotolo, and Lynn Smith-Lovin.
		\newblock Sex and race homogeneity in naturally occurring groups.
		\newblock \emph{Social Forces}, 74\penalty0 (1):\penalty0 15--52, 1995.
		
		\bibitem[Elbaum et~al.(2024)Elbaum, Perry, and
		Messinger]{elbaum2024investigating}
		Batya Elbaum, Lynn~K Perry, and Daniel~S Messinger.
		\newblock Investigating children's interactions in preschool classrooms: An
		overview of research using automated sensing technologies.
		\newblock \emph{Early childhood research quarterly}, 66:\penalty0 147--156,
		2024.
		
		\bibitem[Horn et~al.(2024)Horn, Karsai, and Markova]{horn2024automated}
		Lisa Horn, M{\'a}rton Karsai, and Gabriela Markova.
		\newblock An automated, data-driven approach to children's social dynamics in
		space and time.
		\newblock \emph{Child Development Perspectives}, 18\penalty0 (1):\penalty0
		36--43, 2024.
		
		\bibitem[Santos et~al.(2015)Santos, Daniel, Fernandes, and
		Vaughn]{santos2015affiliative}
		Ant{\'o}nio~J Santos, Joao~R Daniel, Carla Fernandes, and Brian~E Vaughn.
		\newblock Affiliative subgroups in preschool classrooms: Integrating constructs
		and methods from social ethology and sociometric traditions.
		\newblock \emph{PloS one}, 10\penalty0 (7):\penalty0 e0130932, 2015.
		
		\bibitem[Paulus(2018)]{paulus2018preschool}
		Markus Paulus.
		\newblock Preschool children’s and adults’ expectations about interpersonal
		space.
		\newblock \emph{Frontiers in Psychology}, 9:\penalty0 400891, 2018.
		
		\bibitem[Chen et~al.(2019)Chen, Lin, Justice, and Sawyer]{chen2019social}
		Jing Chen, Tzu-Jung Lin, Laura Justice, and Brook Sawyer.
		\newblock The social networks of children with and without disabilities in
		early childhood special education classrooms.
		\newblock \emph{Journal of autism and developmental disorders}, 49:\penalty0
		2779--2794, 2019.
		
		\bibitem[Locke et~al.(2013)Locke, Kasari, Rotheram-Fuller, Kretzmann, and
		Jacobs]{locke2013social}
		Jill Locke, Connie Kasari, Erin Rotheram-Fuller, Mark Kretzmann, and Jeffrey
		Jacobs.
		\newblock Social network changes over the school year among elementary
		school-aged children with and without an autism spectrum disorder.
		\newblock \emph{School Mental Health}, 5:\penalty0 38--47, 2013.
		
		\bibitem[Chamberlain et~al.(2007)Chamberlain, Kasari, and
		Rotheram-Fuller]{chamberlain2007involvement}
		Brandt Chamberlain, Connie Kasari, and Erin Rotheram-Fuller.
		\newblock Involvement or isolation? the social networks of children with autism
		in regular classrooms.
		\newblock \emph{Journal of autism and developmental disorders}, 37:\penalty0
		230--242, 2007.
		
		\bibitem[Dai(2022)]{dai2022thesis}
		Sicheng Dai.
		\newblock \emph{{Study of dynamical social networks of pre-school children
				using wearable wireless sensors}}.
		\newblock Theses, {Universit{\'e} de Lyon ; East China normal university
			(Shanghai)}, May 2022.
		\newblock URL \url{https://theses.hal.science/tel-04010766}.
		
		\bibitem[Leinhardt(1973)]{leinhardt1973development}
		Samuel Leinhardt.
		\newblock The development of transitive structure in children's interpersonal
		relations.
		\newblock \emph{Behavioral science}, 18\penalty0 (4):\penalty0 260--271, 1973.
		
		\bibitem[Dai et~al.(2020)Dai, Bouchet, Nardy, Fleury, Chevrot, and
		Karsai]{dai2020temporal}
		Sicheng Dai, H{\'e}l{\`e}ne Bouchet, Aur{\'e}lie Nardy, Eric Fleury,
		Jean-Pierre Chevrot, and M{\'a}rton Karsai.
		\newblock Temporal social network reconstruction using wireless proximity
		sensors: model selection and consequences.
		\newblock \emph{EPJ Data Sci.}, 9\penalty0 (1):\penalty0 19, 2020.
		\newblock \doi{https://doi.org/10.1140/epjds/s13688-020-00237-8}.
		
		\bibitem[Bastian et~al.(2009)Bastian, Heymann, and Jacomy]{bastian2009gephi}
		Mathieu Bastian, Sebastien Heymann, and Mathieu Jacomy.
		\newblock Gephi: an open source software for exploring and manipulating
		networks.
		\newblock In \emph{Proceedings of the international AAAI conference on web and
			social media}, volume~3, pages 361--362, 2009.
		
		\bibitem[Breiman(2001)]{breiman2001random}
		Leo Breiman.
		\newblock Random forests.
		\newblock \emph{Machine learning}, 45\penalty0 (1):\penalty0 5--32, 2001.
		
		\bibitem[Hastie et~al.(2009)Hastie, Tibshirani, and
		Friedman]{hastie2009elements}
		Trevor Hastie, Robert Tibshirani, and Jerome Friedman.
		\newblock \emph{The elements of statistical learning: data mining, inference,
			and prediction}.
		\newblock Springer Science \& Business Media, 2009.
		
		\bibitem[Veenman(1996)]{veenman1996effects}
		Simon Veenman.
		\newblock Effects of multigrade and multi-age classes reconsidered.
		\newblock \emph{Review of educational research}, 66\penalty0 (3):\penalty0
		323--340, 1996.
		
		\bibitem[Leroy-Audouin and Suchaut(2007)]{leroy2007revisiting}
		Christine Leroy-Audouin and Bruno Suchaut.
		\newblock Revisiting the pedagogical effectiveness of multigrade classes in
		france.
		\newblock \emph{Revue francaise de pedagogie}, 160\penalty0 (3):\penalty0
		103--118, 2007.
		
		\bibitem[Lambiotte et~al.(2019)Lambiotte, Rosvall, and
		Scholtes]{lambiotte2019networks}
		R.~Lambiotte, M.~Rosvall, and I.~Scholtes.
		\newblock From networks to optimal higher-order models of complex systems.
		\newblock \emph{Nat. Phys.}, 2019.
		\newblock \doi{10.1038/s41567-019-0459-y}.
		
		\bibitem[Iacopini et~al.(2019)Iacopini, Petri, Barrat, and
		Latora]{iacopini2019simplicial}
		Iacopo Iacopini, Giovanni Petri, Alain Barrat, and Vito Latora.
		\newblock Simplicial models of social contagion.
		\newblock \emph{Nat. Commun.}, 10:\penalty0 2485, 2019.
		\newblock \doi{https://doi.org/10.1038/s41467-019-10431-6}.
		
		\bibitem[Cencetti et~al.(2021)Cencetti, Battiston, Lepri, and
		Karsai]{cencetti2021temporal}
		Giulia Cencetti, Federico Battiston, Bruno Lepri, and M{\'a}rton Karsai.
		\newblock Temporal properties of higher-order interactions in social networks.
		\newblock \emph{Sci. Rep.}, 11\penalty0 (1):\penalty0 1--10, 2021.
		\newblock \doi{https://doi.org/10.1038/s41598-021-86469-8}.
		
		\bibitem[Mancastroppa et~al.(2023)Mancastroppa, Iacopini, Petri, and
		Barrat]{mancastroppa2023hyper}
		Marco Mancastroppa, Iacopo Iacopini, Giovanni Petri, and Alain Barrat.
		\newblock Hyper-cores promote localization and efficient seeding in
		higher-order processes.
		\newblock \emph{Nat. Commun.}, 14:\penalty0 6223, 2023.
		\newblock \doi{https://doi.org/10.1038/s41467-023-41887-2}.
		
		\bibitem[Mancastroppa et~al.(2024)Mancastroppa, Iacopini, Petri, and
		Barrat]{mancastroppa2024structural}
		Marco Mancastroppa, Iacopo Iacopini, Giovanni Petri, and Alain Barrat.
		\newblock The structural evolution of temporal hypergraphs through the lens of
		hyper-cores.
		\newblock \emph{arXiv preprint arXiv:2402.06485}, 2024.
		
		\bibitem[Liu et~al.(2020)Liu, Yuan, Lin, Qin, Zhang, and
		Zhou]{liu2020efficient}
		Boge Liu, Long Yuan, Xuemin Lin, Lu~Qin, Wenjie Zhang, and Jingren Zhou.
		\newblock Efficient ($\alpha$, $\beta$)-core computation in bipartite graphs.
		\newblock \emph{The VLDB Journal}, 29\penalty0 (5):\penalty0 1075--1099, 2020.
		
		\bibitem[Lotito et~al.(2022)Lotito, Musciotto, Montresor, and
		Battiston]{lotito2022higher}
		Quintino~Francesco Lotito, Federico Musciotto, Alberto Montresor, and Federico
		Battiston.
		\newblock Higher-order motif analysis in hypergraphs.
		\newblock \emph{Communications Physics}, 5\penalty0 (1):\penalty0 79, 2022.
		
		\bibitem[Suchaut(2010)]{suchaut2010efficacite}
		Bruno Suchaut.
		\newblock Efficacit{\'e} p{\'e}dagogique des classes {\`a} cours double {\`a}
		l’{\'e}cole primaire: le cas du cours pr{\'e}paratoire.
		\newblock \emph{Revue fran{\c{c}}aise de p{\'e}dagogie. Recherches en
			{\'e}ducation}, \penalty0 (173):\penalty0 51--66, 2010.
		
		\bibitem[Veldt et~al.(2023)Veldt, Benson, and
		Kleinberg]{veldt2023combinatorial}
		Nate Veldt, Austin~R Benson, and Jon Kleinberg.
		\newblock Combinatorial characterizations and impossibilities for higher-order
		homophily.
		\newblock \emph{Science Advances}, 9\penalty0 (1):\penalty0 eabq3200, 2023.
		
		\bibitem[Altenburger and Ugander(2018)]{altenburger2018monophily}
		Kristen~M Altenburger and Johan Ugander.
		\newblock Monophily in social networks introduces similarity among
		friends-of-friends.
		\newblock \emph{Nature human behaviour}, 2\penalty0 (4):\penalty0 284--290,
		2018.
		
		\bibitem[Gallo et~al.(2024{\natexlab{b}})Gallo, Zappal{\`a}, Karimi, and
		Battiston]{gallo2024higher_f2f}
		Luca Gallo, Chiara Zappal{\`a}, Fariba Karimi, and Federico Battiston.
		\newblock Higher-order modeling of face-to-face interactions.
		\newblock \emph{arXiv preprint arXiv:2406.05026}, 2024{\natexlab{b}}.
		
		\bibitem[Landry et~al.(2023)Landry, Lucas, Iacopini, Petri, Schwarze, Patania,
		and Torres]{landry2023xgi}
		Nicholas~W Landry, Maxime Lucas, Iacopo Iacopini, Giovanni Petri, Alice
		Schwarze, Alice Patania, and Leo Torres.
		\newblock Xgi: A python package for higher-order interaction networks.
		\newblock \emph{Journal of Open Source Software}, 8\penalty0 (85):\penalty0
		5162, 2023.
		
	\end{thebibliography}

\end{document}